\documentclass[reprint,amsmath,amssymb,aps,]{revtex4-1}

\usepackage{amsmath}
\usepackage{graphicx}% Include figure files
\usepackage{dcolumn}
\usepackage{bm}
\usepackage[normalem]{ulem}
\usepackage[colorlinks=true,linkcolor=blue,urlcolor=blue,citecolor=blue]{hyperref}
\usepackage{soul}

\begin{document}

\title{Observation of optical Rabi oscillations in transmission signal of atomic vapor under continuous-wave laser excitation}

\author{Aram Papoyan}
\author{Svetlana Shmavonyan}
\affiliation{Institute for Physical Research, National Academy of Sciences of Armenia, 0203 Ashtarak-2, Armenia}

\begin{abstract}
We have studied the temporal behavior of the atomic absorption signal under resonant excitation with a continuous-wave laser radiation. Measurements done for D$_2$ line of $^{85}$Rb with $\approx$ 1 ns temporal resolution have shown irregular oscillatory behavior of the transmission signal, which becomes well pronounced for high laser power, and disappears when the laser is tuned off-resonance. Application of the fast Fourier transform analysis of the transmission signal reveals power-dependent frequency peaks, which are shown to be associated with Rabi frequency. Possible linkage of the observed results with the phase-to-amplitude noise conversion caused by the the phase fluctuations of laser field is discussed.
\end{abstract}
\maketitle

\section{Introduction}
\label{intro}

Resonant interaction of narrow-linewidth cw laser radiation with atomic vapor (notably, alkali metal vapor) is intensely studied in the past decades, driven by fundamental interest and important emerging applications. Most of these studies deal with a steady-state regime of interaction of atomic ensemble with resonant light required for establishment of the relevant processes. The steady-state regime implies an onset of a dynamic balance between elementary processes (e.g. absorption and emission), which leads to the invariance of the average level of an atomic signal in time under invariable excitation conditions. But a question remains: is the stationary signal time-independent in a short time scale? The dynamic contribution of elementary processes (Rabi cycle) can be revealed in measurements with high temporal resolution.

Rabi oscillations were directly observed in microwave domain in many experiments, meanwhile this task becomes much more problematic in the optical frequency range, notably for the transitions with large dipole moment. Optical Rabi oscillations with a half period as short as 1 ns have been detected in single nitrogen vacancy centers in diamond \cite{robledo}. Direct detection of the time-resolved Rabi oscillations in the coherent transient response of the resonance fluorescence of a single charged InAs quantum dot has been reported in \cite{schaibley}. Ultrafast Rabi oscillation of an atom ensemble in Gaussian spatial distribution has been investigated in \cite{lee} using the cold atomic rubidium vapor spatially confined in a magneto-optical trap. Rabi oscillations in the spatial profiles of yoked superfluorescent pulses from rubidium vapor have been observed in \cite{kitano} by driving the 5S -- 5D two-photon transition with a femtosecond laser. All these experiments were done with pulsed laser excitation. A nanosecond-range spiking dynamics of frequency up-converted emission generated on the 6P$_{3/2}$ -- 5S$_{1/2}$ transition in Rb vapor under two-photon excitation of the 5D$_{5/2}$ level with bichromatic cw laser light is reported in \cite{akulshin}. The spikes were attributed to temporal properties of the directional generation on the 5D$_{5/2}$ -- 6P$_{3/2}$ transition, and explained by the quantum-mechanical nature of the cooperative effects.

As is known since thee decades, intrinsic stochastic phase fluctuations in cw laser (notably, diode laser) field, which determine the finite linewidth of the laser radiation, are converted to amplitude modulation during the resonant interaction with atomic systems \cite{haslwanter,ritsch,anderson}.  Theory of this process has been developed in \cite{anderson,walser,frueholz}, showing good consistency with experimental results. The phenomenon of phase-to-amplitude noise conversion underlies spectroscopic technique developed in early 90-s: illuminating the atomic medium with broadband laser radiation with fixed central frequency and performing spectral decomposition of noisy atomic response, one can reveal resonant spectral features \cite{yabuzaki,mcintyre,walser}. 
The theory of phase-to-amplitude noise conversion predicts two distinct temporal components of the atom's response: an adiabatic component and a nonadiabatic component manifesting itself as population variations oscillating at the Rabi frequency \cite{frueholz}. The latter was experimentally verified in \cite{camparo1}, where authors found that the atoms behave like damped, driven harmonic oscillators, and the Fourier spectrum of the atom's population fluctuations shows a resonance at the Rabi frequency on the microwave transition coupling hyperfine sublevels of $^{87}$Rb, which redistributes population depletion imposed by an optical pumping. Later, the studies were done also for the optically thick vapor \cite{camparo2}. The role of laser linewidth in conversion of laser phase noise to amplitude noise in a resonant atomic vapor was analized in \cite{camparo3}. Different behavior of optical spin noise spectra in Rb vapor for homogeneous and inhomogeneous broadenings was demonstrated in \cite{ma}. Applications of spin noise spectroscopy for studies of magnetic resonances, linear optics, and Raman scattering were addressed in \cite{zapasskii,glazov,swar}. The same technique was used in \cite{katsoprinakis} for measurement of transverse spin-relaxation rates in a rubidium vapor. In \cite{kitching} it was shown that optical pumping plays a significant role in determining the noise of laser-pumped vapor-cell microwave frequency standards. Noise spectroscopy was applied in \cite{martinelli,valente} to study nonlinear magneto-optical resonances in Rb vapor. Experimental observation of quantum noise in the polarization of laser light passing through a dense Rb vapor using a homodyne detection technique was reported in \cite{zibrov}.

In this paper we report on the temporal structure in atomic absorption signal under continuous-wave laser excitation observed with fast detection and processing technique. The experiment is done for D$_2$ line of rubidium in a simple configuration with atomic vapor cell.

\section{Experimental arrangement}
\label{exp}

\begin{figure}[h!]
	\centering
	\begin{center}
		\includegraphics[width=220pt]{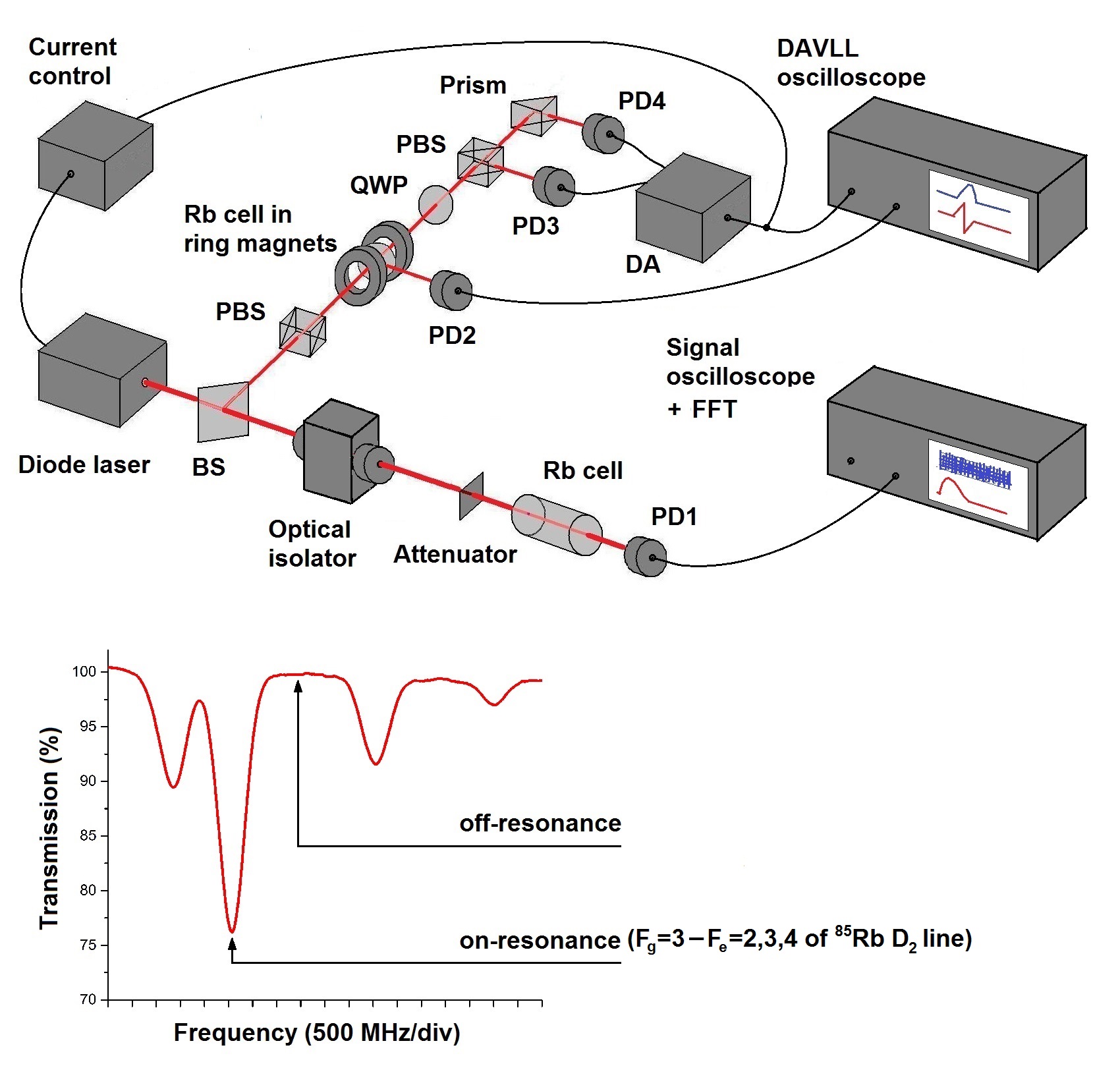}
		% Here is how to import EPS art
		\caption{\label{fig:Fig1} Upper panel: schematic drawing of the experimental setup. BS -- beam splitter; PBS -- polarizing beam splitter; QWP -- quarter-wave plate; PD1 - PD4 -- photodiodes; DA -- differential amplifier. Lower panel: laser radiation frequency positions across $^{85}$Rb and $^{87}$Rb D$_2$ line explored in the measurements (the spectrum is recorded for the laser power of 0.5 mW).}
	\end{center}
\end{figure}

The sketch of experimental setup is shown in the upper panel of Fig. \ref{fig:Fig1}. It comprised of a simple main arrangement for dynamic recording of the radiation transmitted through a rubidium vapor cell, and a more complex auxiliary arrangement for tuneable locking of laser radiation frequency.

An unfocused $\approx$ 2-mm-diameter linearly-polarized beam from a single-frequency free-running cw Fabry-P\'erot diode laser (wavelength 780 nm, light power 30 mW, linewidth $\approx$ 20 MHz) was directed into a 70 mm-long, 10 mm-diameter sealed cylindrical cell with a side arm containing natural rubidium. An optical isolator was used to prevent laser cavity effect imposed by reflections; the power of laser radiation entering the cell was controlled by a circular variable neutral density filter. The vapor-cell temperature was kept at 27.8$^{\circ}$C, which corresponds to a number density of Rb atoms $N$ = 9.27$\times10^{9}$ cm$^{-3}$. The radiation transmitted through the cell was detected by a Newport 818-BB-21A optical biased receiver (PD1) with 1.2 GHz bandwidth and 0.5 ns rise/fall time. This photodetector with a sensor diameter of 0.4 mm was mounted on a translation stage enabling adjustment of its position across the transmitted beam aperture.

Signal from the photodetector PD1 was monitored by the Tektronix TDS3032B digital storage oscilloscope with 300 MHz bandwidth and 2.5 GS/s sample rate per channel. An integrated fast Fourier transform (FFT) module TDS3FFT built in the oscilloscope allowed to perform a real-time Fourier analysis of the photodetector signal. Another photodetector, facing the cell wall (not shown in Fig. \ref{fig:Fig1}) was installed to monitor the fluorescence signal with separate oscilloscope.

Small fraction of the laser beam was branched onto a dichroic atomic vapor laser lock (DAVLL) scheme \cite{yashchuk} allowing realization of locking the laser radiation frequency to atomic resonance line employing linear magneto-optics. The laser frequency was locked to the chosen atomic resonance via a feedback loop controlling the injection current of laser diode, with a dispersion-shaped error signal formed by the differential amplifier. 

The following procedure was employed for the measurements. First, the laser was locked to the Doppler-overlapped atomic transition group $^{85}$Rb $F_g$=3 $\to$ $F_e$=2,3,4 of the Rb D$_2$ line (see lower panel of Fig. \ref{fig:Fig1}), and absorption (PD1) signal was recorded for a 2 $\mu$s interval (no averaging, 10000 measurement points with the step of 0.2 ns). The measurement data were stored together with the real-time Fourier transform spectrum generated by an FFT module operating in linear root-mean-square output and rectangular window mode, which assures maximum frequency accuracy. Afterwards, the same measurements were done with the laser radiation frequency tuned to the off-resonance position. 

The measurements were done for 8 values of the laser radiation power $P_L$: 2, 5, 8, 11, 14, 17, 20, and 23 mW. We should note that the real power entering the cell was 1.18 times less because of reflection from the sapphire window cut across the c-axis to avoid birefringence. The net absorption of the transmitted beam at atomic resonance ranged from 30.5\% for the lowest power to 9.55\% for the highest power. Throughout the measurements, the PD1 photodetector sensor was carefully aligned to the center of the transmitted beam.

\section{Results and discussion}
\label{results} 

\begin{figure}[h!]
	\centering
	\begin{center}
		\includegraphics[width=240pt]{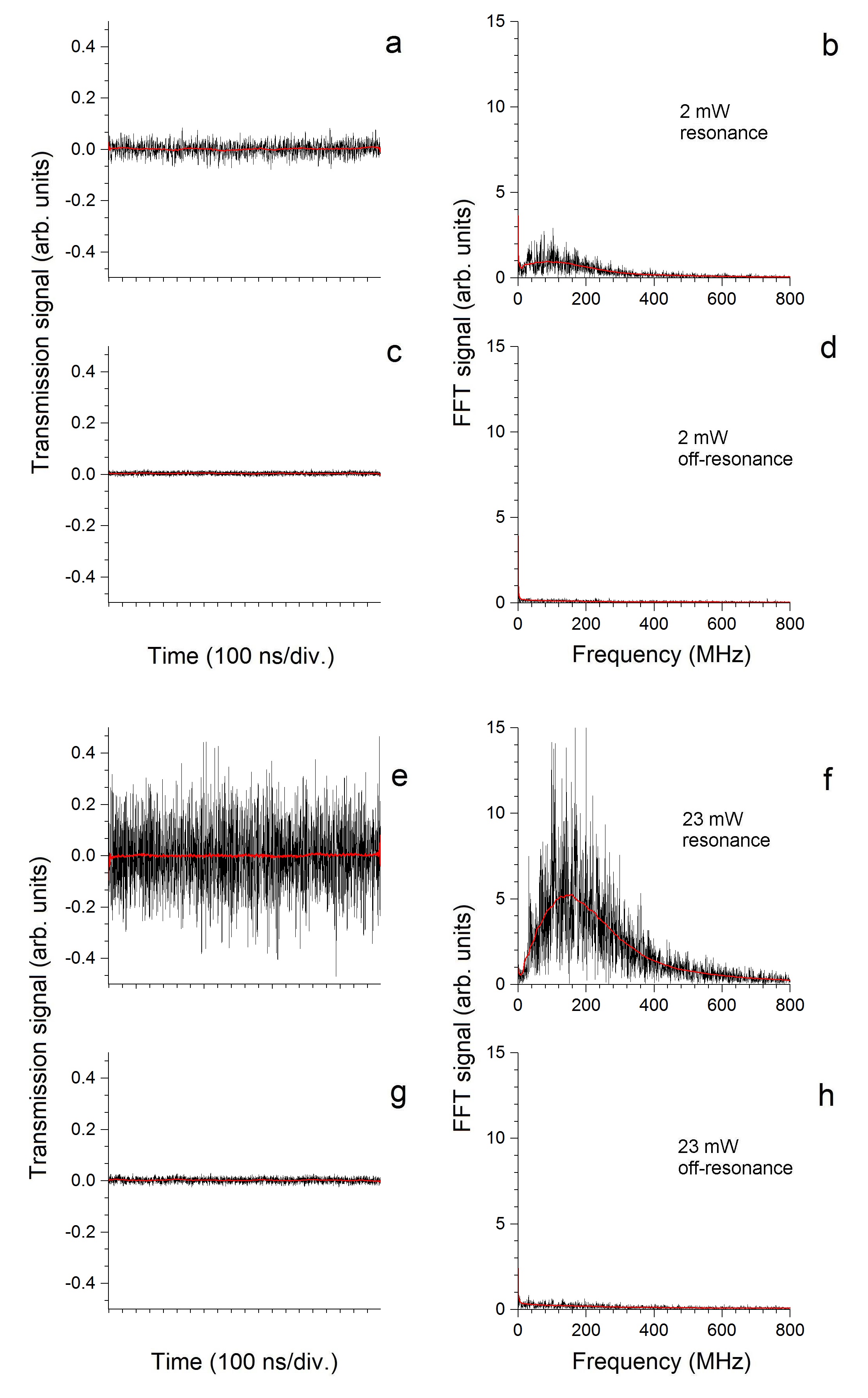}
		% Here is how to import EPS art
		\caption{\label{fig:Fig2} Directly recorded resonance and off-resonance transmission signals (a,c,e,g) and corresponding FFT spectra (b,d,f,h) for incident laser power 2 mW (a-d) and 23 mW (e-h). The same amplitude scales are used for all the transmission (left column) and FFT (right column) graphs. Red curves show the post-measurement adjacent averaging with 500 points of window.}
	\end{center}
\end{figure}

The measurements results for $P_L$ = 2 and 23 mW are presented in Fig. \ref{fig:Fig2}. One can clearly see an irregular oscillatory behavior of the transmission signal when the laser radiation frequency is tuned to the atomic resonance (graphs $a$ and $e$), with the magnitude increasing with $P_L$. Noteworthy, the temporal structure gets outright vanished already at integration time of 100 ns (see red lines in graphs $a$ and $e$). The oscillations practically disappear when the laser frequency is driven out of the resonance (graphs $c$ and $g$). Application of a fast Fourier transform (right column graphs) allows revealing the frequency spectrum of the signal, which exhibits a peak behavior, with a maximum frequency rising with $P_L$.

\begin{figure}[h!]
	\centering
	\begin{center}
		\includegraphics[width=200pt]{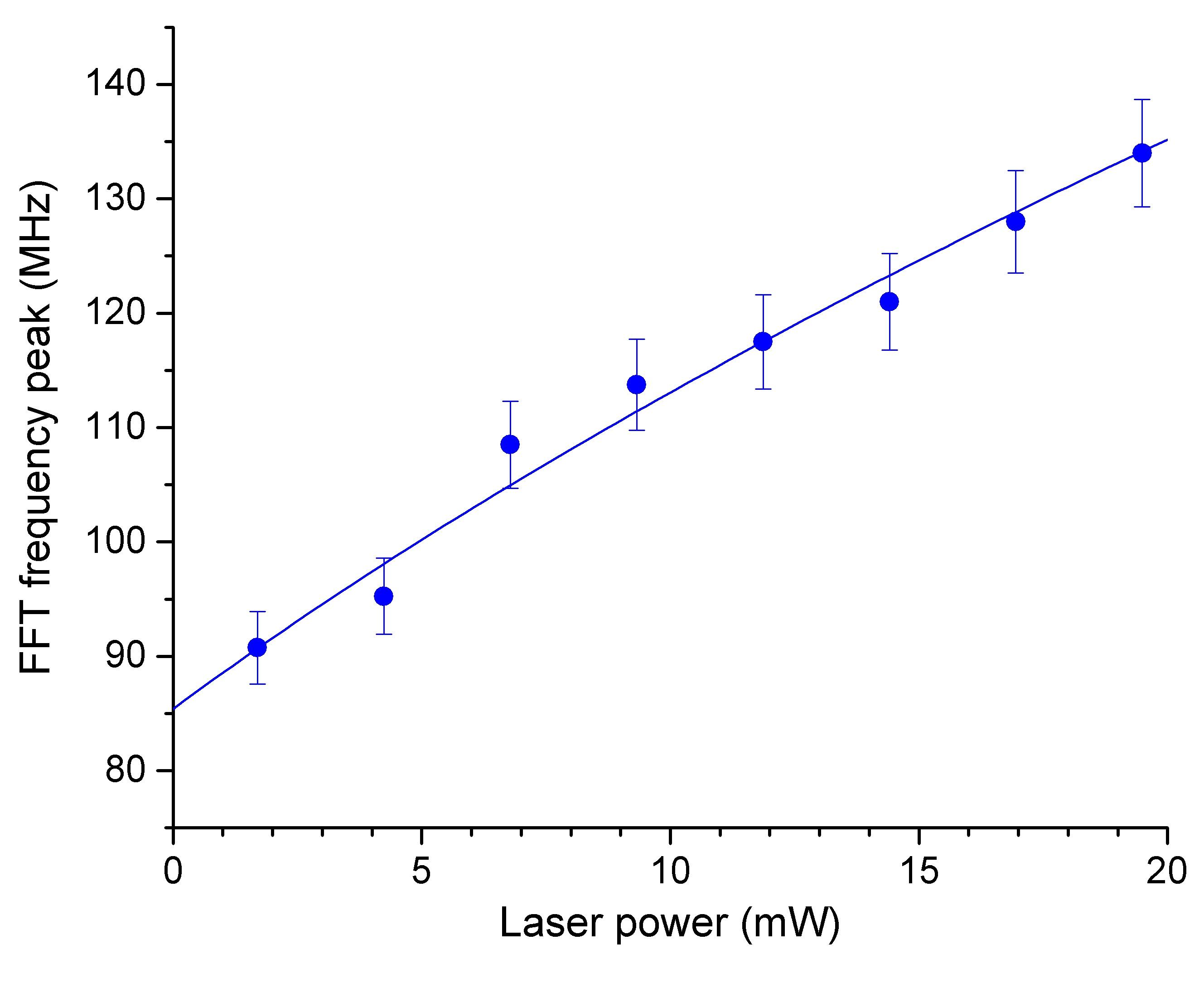}
		% Here is how to import EPS art
		\caption{\label{fig:Fig3} Dependence of the peak frequency of FFT signal on the laser power entering the cell. Solid dots with error bars: experiment; solid line: calculated with Eq. \ref{eq::effectrabi}.}
	\end{center}
\end{figure}

Experimental dependence of the peak frequency of the FFT signal on the incident laser radiation power is plotted in Fig. \ref{fig:Fig3}. Solid dots present the averaged values of 5 measurements for each value of $P_L$, with an error of $\pm$3.5\%. The square-root-shaped course of this dependence allows us to suppose that the oscillatory behavior of the transmission signal should be attributed to dynamic Rabi cycle. To verify this supposition, we have made estimates for the expected generalized (effective) Rabi frequency $\widetilde{\Omega}$:

\begin{equation}
\label{eq::effectrabi}
\widetilde{\Omega}_{ij}=\sqrt{\frac{d^2_{ij}P_L}{\epsilon_0 nc \hbar^2 S}+\Delta^2}.
\end{equation}

\noindent (see e.g. \cite{aleksanyan}), where $d_{ij}$ is the transition dipole moment, $\epsilon_0$ is the vacuum permittivity, $n$ is the refractive index, $c$ is the speed of light, $S$ is the effective area of the laser beam cross-section, and $\Delta$ is the detuning of the laser radiation frequency from the atomic transition (including also Doppler broadening and the spectral linewidth of laser radiation when it exceeds transition natural linewidth).

The main contribution in the Doppler-overlapped ($\approx$ 640 MHz FWHM) hyperfine transition group $^{85}$Rb $F_g$=3 $\to$ $F_e$=2,3,4 gives the cycling transition $F_g$=3 $\to$ $F_e$=4, having a strength factor, which exceeds the strength factors of $F_g$=3 $\to$ $F_e$=3 and $F_g$=3 $\to$ $F_e$=2 by factors of 2.3 and 8.1, respectively. As a result, also the frequency of this transition is the closest to the transmission minimum ($\approx$ 40 MHz). For this reason, in calculations we have used for $d_{ij}$ the effective dipole moment 1.659$\times$10$^{-29}$ C/m of the $F_g$=3 $\to$ $F_e$=4 transition \cite{steck}. The calculated dependence is plotted in Fig. \ref{fig:Fig3} by a solid line. The best fit has been obtained with the effective detuning $\Delta$ = 85.4 MHz, which is realistic given the experimental uncertainties (contributions from other transitions, laser beam profile, thermal motion of atoms across the beam, Doppler broadening, laser linewidth, relaxations, etc.).

Moreover, it is important to note that the frequency corresponding to the maximum of FFT signal does not correspond to the maximum of Rabi frequency. Analyzing the shape of FFT spectrum, one can say that the most intense interaction occurs with the velocity group of atoms, which are in exact resonance with the laser field, and these atoms contribute to the far high-frequency wing of the FFT profile. The most intense FFT signal corresponds to the contribution of atoms with the mean longitudinal thermal velocity. One may say that the FFT profile somehow reflects Maxwellian velocity distribution. 

\begin{figure}[h!]
	\centering
	\begin{center}
		\includegraphics[width=180pt]{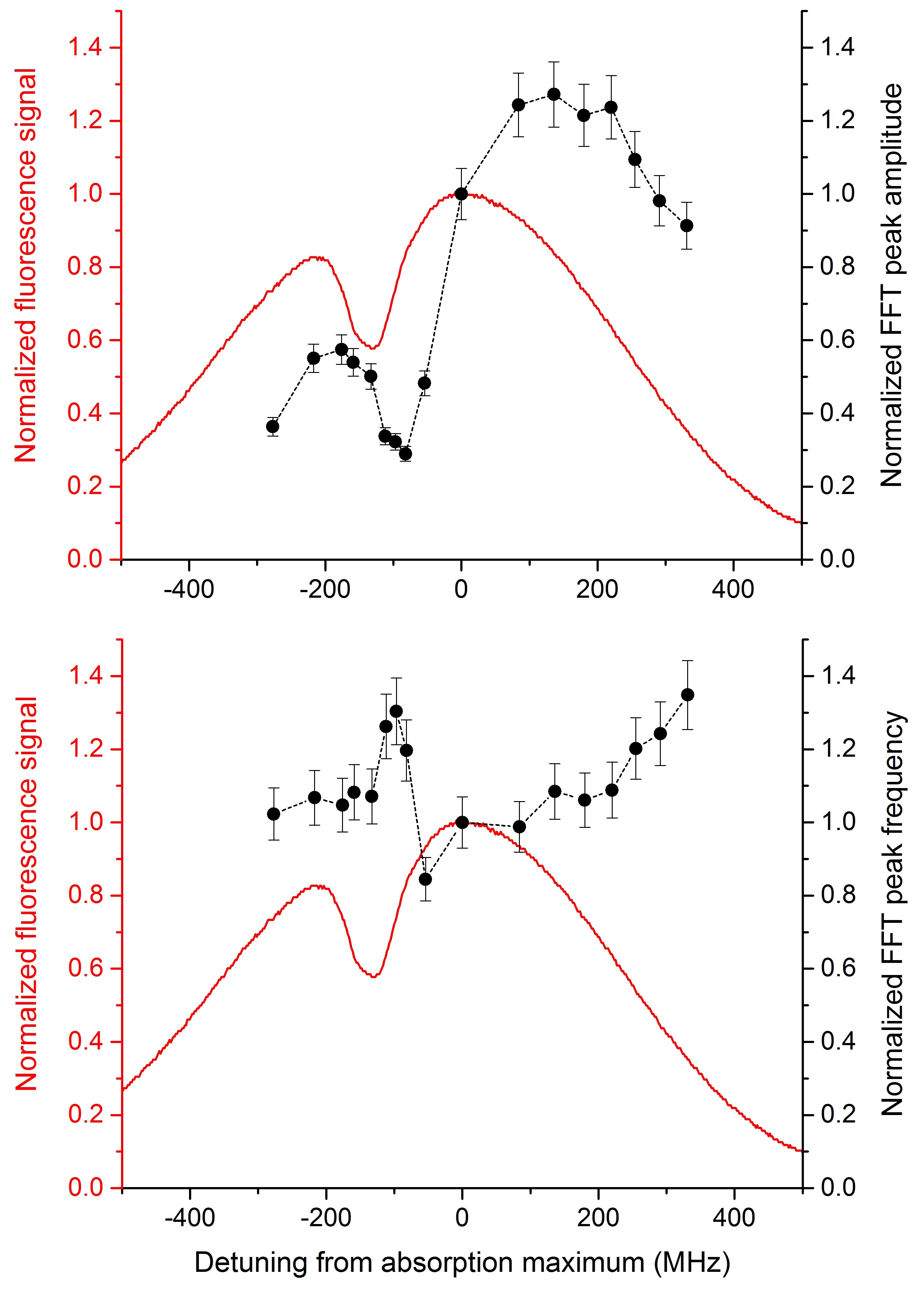}
		% Here is how to import EPS art
		\caption{\label{fig:Fig4} Black solid dots: dependences of FFT signal peak amplitude  (upper panel) and frequency (lower panel) on detuning of laser frequency from the center of Doppler-overlapped $F_g$=3 $\to$ $F_e$=2,3,4 transition group for the incident laser power 23 mW/cm$^2$. Dependences are normalized to zero-detuning values (corresponding fluorescence spectrum is presented by red solid line).}
	\end{center}
\end{figure}

In the next run of measurements we have studied the dependence of FFT spectrum on laser frequency across the Doppler-overlapped $F_g$=3 $\to$ $F_e$=2,3,4 transition group for the highest value 23 mW/cm$^2$ of incident laser power. Fig. \ref{fig:Fig4} presents the amplitude and frequency of FFT spectrum versus detuning of laser frequency from the maximum absorption position (17 values, ranged from -277 to +331 MHz). The obtained results are presented along with the fluorescence spectrum recorded with slow scanning (steady-state interaction regime) in the same experimental conditions. Note that although the cell windows were tilted from the light propagation axis to prevent overlapping of the forward and reflected beams, a sub-Doppler dip appears in fluorescence spectrum at the frequency position of two overlapped crossover resonances linked with the cycling transition $F_g$=3 $\to$ $F_e$=4. The study of origin of this feature is out of the scope of this paper, and will be discussed elsewhere. 

As can be seen from the upper panel of Fig. \ref{fig:Fig4}, the dependence of the FFT signal amplitude on the laser frequency qualitatively repeats the spectrum of the atomic signal, with a variable positive frequency shift and an asymmetric amplitude behavior (lower amplitudes on the negative wing, and higher amplitudes on the positive one). Noteworthy, the same dependence is obtained also for the amplitude of fluctuations of transmission signal directly recorded by a photodetector  (not shown in Fig. \ref{fig:Fig4}). Supposing that the our results are linked with the phase-to-amplitude noise conversion mechanism, one should expect a symmetric double-peaked structure for the atomic signal fluctuations, with the minimum noise at the resonance frequency (see e.g. \cite{anderson,kitching}). When the laser is broadband, covering several transitions within the Doppler profile, the structure becomes asymmetric, as the resonance profile is diminished on the side where the noncycling transitions occur \cite{mcintyre,ma}. We should note that although in our experiment the laser linewidth exceeds the natural linewidth by only $\approx$ 3.5 times, we do not observe a pronounced double-peaked structure. So, the observed asymmetry should be mostly attributed to correlated nature of the signal mentioned in \cite{mcintyre}.

Different behavior is observed for the FFT peak frequency versus the laser frequency detuning (see the lower panel in Fig. \ref{fig:Fig4}). Here, the peak frequency continues to grow on the blue wing of the Doppler-broadened resonance, which, we believe, is due to the growing contribution in the overall signal of the strongest transition $F_g$=3 $\to$ $F_e$=4 having the highest frequency. 

\begin{figure}[h!]
	\centering
	\begin{center}
		\includegraphics[width=180pt]{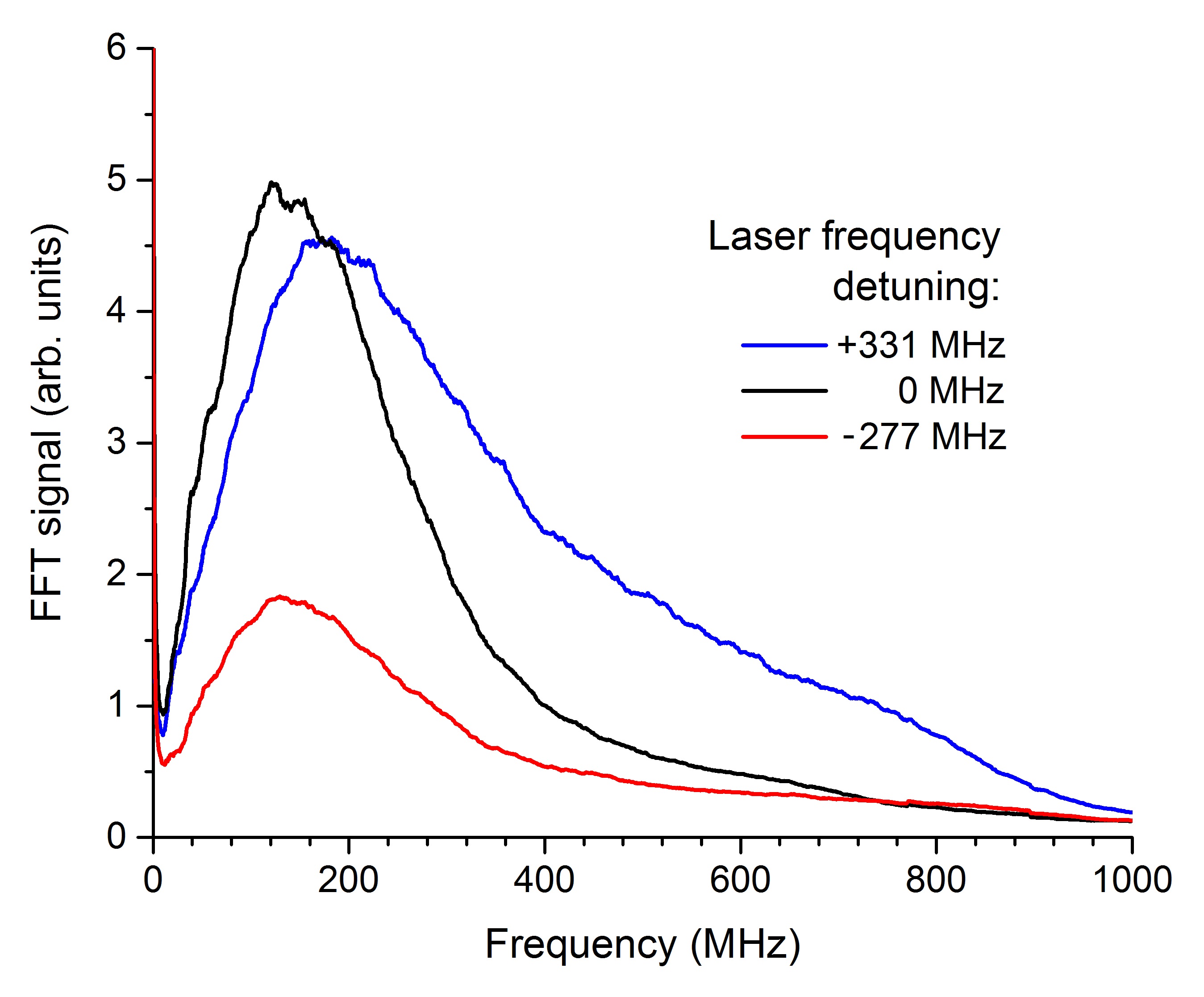}
		% Here is how to import EPS art
		\caption{\label{fig:Fig5} FFT spectra recorded with the incident laser power 23 mW/cm$^2$ for 3 values of detuning of laser radiation frequency from the center of Doppler-overlapped $F_g$=3 $\to$ $F_e$=2,3,4 transition group.}
	\end{center}
\end{figure}

Detuning of the laser frequency from the Doppler-overlapped resonance affects also the shape of the FFT spectrum (see Fig. \ref{fig:Fig5}). The FFT resonance becomes broader as the laser frequency is detuned from the center of the Doppler-overlapped $F_g$=3 $\to$ $F_e$=2,3,4 transition group.

\begin{figure}[h!]
	\centering
	\begin{center}
		\includegraphics[width=170pt]{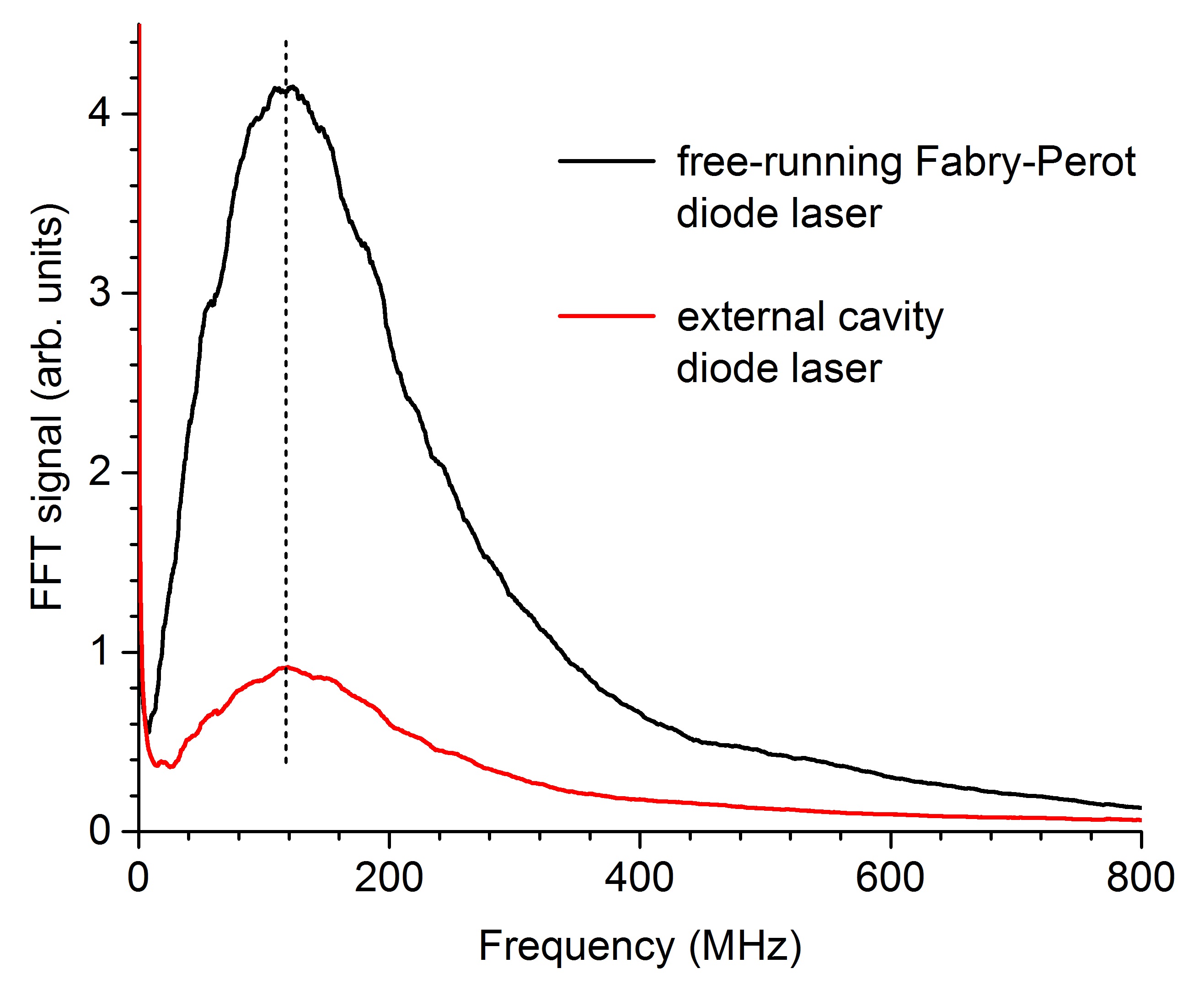}
		% Here is how to import EPS art
		\caption{\label{fig:Fig6} FFT spectra of transmission signals for the free-running Fabry-P\'erot (black line) and external cavity (red line) diode lasers with the same incident intensity $\approx$ 540 mW/cm$^2$.}
	\end{center}
\end{figure}

In order to clarify the role of laser radiation linewidth, additional measurements similar to those presented in Fig. \ref{fig:Fig2} have been done by replacing the free running Fabry-P\'erot diode laser with a single-frequency cw external cavity diode laser ECDL-7850R (Atrix Management S.A.) having 30 mW radiation power and 1 MHz linewidth, and otherwise invariable experimental conditions. Fig. \ref{fig:Fig6} combines 500-point averaged FFT spectra of transmission signal measured with lasers having spectral linewidths $\Delta \nu_L \approx$ 20 MHz (black line) and 1 MHz (red line) for the same radiation intensity. To obtain the same intensity with two lasers, a 1 mm pinhole was installed in front of the cell, centered along the beam axis, and the attenuator provided the same power for the two lasers. As is clearly seen, laser linewidth affects the magnitude of FFT signal, while the profile in general and peak position in particular remain practically invariable, with the scaling factor of 4.6. This result is consistent with findings of \cite{camparo3}, where it was shown that for single-mode laser linewidths less than the atomic dephasing rate, the signal fluctuations, and hence the FFT amplitude, depend on laser linewidth as $\sqrt{\Delta \nu_L}$.

Finally, let us comment on the ``permanence" of the observed Rabi oscillations for continuous-wave excitation of atomic vapor. Here we should note that each particular atom interacts with the laser field within a limited time (a $\mu$s-range time of flight of an atom through the laser beam). As a result, at each moment of time the overall dynamic equilibrium of atoms in the laser beam is maintained by a rapid succession of departing and arriving atoms, resembling the pulsed interaction regime. 

\section{Conclusions and outlook}
\label{conclusion}

In conclusion, we have observed and qualitatively analyzed the appearance of a temporal structure in the atomic vapor resonant transmission signal. This observation became possible thanks to usage of an ultrafast detection system with small-aperture photodetector, and application of a fast Fourier transform signal processing. Based on the laser power dependence of the FFT spectrum peak, we have attributed the observed fluctuations to the optical Rabi oscillations in an ensemble of Rb atoms composing a thermal vapor.

Experimental studies of spectral and laser-linewidth dependences of the FFT signal have been done in order to check the possible role of the extensively studied phenomenon of conversion of laser field phase noise to atomic signal amplitude noise (the phase diffusion and jump models) in the observed features. Some of our experimental results are qualitatively consistent with the phase-to-amplitude noise conversion mechanism (notably, the red-blue wings asymmetry in the laser frequency dependence of the FFT signal amplitude and the square-root dependence of the FFT signal amplitude on the laser linewidth). However, in our experiment we have not observed a pronounced double-peaked structure inherent for the phase-to-amplitude conversion process, especially expected for the strong-field saturation regime explored in our case.

Measurements done with two different diode lasers (a free-running Fabry-P\'erot and an external cavity, linewidths 20 MHz and 1 MHz, respectively) have shown different amplitudes of the FFT signal for the same intensity. However, the peak position and the shape of FFT spectrum profile remain invariable. This indicates that the measured optical Rabi frequency is independent of the phase noise mechanism for the laser field.

As a follow-up action, we plan to perform a comprehensive and detailed experimental study combined with rigorous theoretical modeling, which will properly account for all the physical processes involved. The theoretical study should be primarily focused on the main challenge in direct observation of Rabi flopping, that is justification of synchronization between $\sim$ 10$^9$ different atoms contributing to the signal, taking into account their thermal motion and random phases, which can completely wash out the effect. Further experiments are planned to be done, in particular, using optical nanocells, where the number of contributing atoms in the interaction region is orders of magnitude less, and the Doppler broadening is strongly suppressed (see, e.g. \cite{sarkisyan}). 

From the practical point of view, the obtained results can be used for characterization of both the laser field (particularly, determination of spectral linewidth and spatial distribution of intensity across the beam) and atomic vapor cells (the shape and width of the FFT spectrum profile carries information about the optical length of the cell, the vapor density, the presence of buffer gas or antirelaxation coating, etc.). 

\textbf{Acknowledgments.} The authors are grateful to D. Sarkisyan and D. Budker for stimulating discussions, and to I. Novikova for drawing our attention to the phase-to-amplitude noise conversion mechanism. This work was supported by the State Committee of Science MES RA, in frame of the research project No.18T-1C234.


\begin{thebibliography}{1}
	
\bibitem{robledo} L. Robledo, H. Bernien, I. van Weperen, R. Hanson, ``Control and coherence of the optical transition of single nitrogen vacancy centers in diamond", Phys. Rev. Lett. \textbf{105}, 177403 (2010).

\bibitem{schaibley} J.R. Schaibley, A.P. Burgers, G.A. McCracken, D.G. Steel, A.S. Bracker, D. Gammon, L.J. Sham, ``Direct detection of time-resolved Rabi oscillations in a single quantum dot via resonance fluorescence", Phys. Rev. B \textbf{87}, 115311 (2013).

\bibitem{lee} H-G. Lee, H. Kim, J. Ahn, ``Ultrafast laser-driven Rabi oscillations of a trapped atomic vapor", Opt. Lett. \textbf{40}, 510 (2015).

\bibitem{kitano} K. Kitano, H. Maeda, ``Rabi oscillations in the spatial profiles of superfluorescent pulses from rubidium vapor", Opt. Express \textbf{25}, 23826 (2017).

\bibitem{akulshin} A. Akulshin, N. Rahaman, S. Suslov, D. Budker, R. McLean, ``Spiking dynamics of frequency up-converted field generated in continuous-wave excited rubidium vapours", J. Opt. Soc. Am. B (accepted); \url{arXiv:2003.06149} [physics.optics].

\bibitem{haslwanter} Th. Haslwanter, H. Ritsch, J. Cooper, P. Zoller, ``Laser-noise-induced population fluctuations in two- and three-level systems", Phys. Rev. A \textbf{38}, 5652 (1988).

\bibitem{ritsch} H. Ritsch, P. Zoller, J. Cooper, ``Power spectra and variance of laser-noise-induced population fluctuations in two-level atoms", Phys. Rev. A \textbf{41}, 2653 (1990).

\bibitem{anderson} M.H. Anderson, R.D. Jones, J. Cooper, S.J. Smith, D.S. Elliott, H. Ritsch, P. Zoller, ``Variance and spectra of fluorescence-intensity fluctuations from two-level atoms in a phase-diffusing field", Phys. Rev. A \textbf{42}, 6690 (1990).

\bibitem{walser} R. Walser, P. Zoller, ``Laser-noise-induced polarization fluctuations as a spectroscopic tool", Phys. Rev. A \textbf{49}, 5067 (1994).

\bibitem{frueholz} R.P. Frueholz, J.C. Camparo, ``Underlying simplicity of atomic population variations induced by a stochastic phase-fluctuating field", Phys. Rev. A 54, 3499 (1996).

\bibitem{yabuzaki} T. Yabuzaki, T. Mitsui, U. Tanaka, ``New type of high-resolution spectroscopy with a diode laser", Phys. Rev. Lett. \textbf{67}, 2453 (1991).

\bibitem{mcintyre} D.H. McIntyre, C.E. Fairchild, J. Cooper, R. Walser, ``Diode-laser noise spectroscopy of rubidium", Opt. Lett. \textbf{18}, 1816 (1993).

\bibitem{camparo1} J.C. Camparo, J.G. Coffer, R.P. Frueholz, ``Temporal response of an atom to a stochastic field: Resonant enhancement of population fluctuations at the Rabi frequency", Phys. Rev. A \textbf{56}, 1007 (1997).

\bibitem{camparo2} J.C. Camparo, ``Conversion of laser phase noise to amplitude noise in an optically thick vapor", J. Opt. Soc. Am. B \textbf{15}, 1177 (1998).

\bibitem{camparo3} J.C. Camparo, J.G. Coffer, ``Conversion of laser phase noise to amplitude noise in a resonant atomic vapor: The role of laser linewidth", Phys. Rev. A \textbf{59}, 728 (1999).

\bibitem{ma} J. Ma, P. Shi, X. Qian, Y. Shang, Y. Ji, ``Optical spin noise spectra of Rb atomic gas with homogeneous and inhomogeneous broadening", Sci. Rep. \textbf{7}, 10238 (2017).

\bibitem{zapasskii} V.S. Zapasskii, ``Spin-noise spectroscopy: from proof of principle to applications", Advances in Optics and Photonics \textbf{5}, 131 (2013).

\bibitem{glazov} M.M. Glazov, V.S. Zapasskii, ``Linear optics, Raman scattering, and spin noise spectroscopy", Opt. Express \textbf{23}, 11713 (2015).

\bibitem{swar} M. Swar, D. Roy, D. Dhanalakshmi, S. Chaudhuri, S. Roy, H. Ramachandran, ``Measurements of spin properties of atomic systems in and out of equilibrium via noise spectroscopy", Opt. Express \textbf{26}, 32168 (2018).

\bibitem{katsoprinakis} G.E. Katsoprinakis, A.T. Dellis, I.K. Kominis, ``Measurement of transverse spin-relaxation rates in a rubidium vapor by use of spin-noise spectroscopy", Phys. Rev. A \textbf{75}, 042502 (2007).

\bibitem{kitching} J. Kitching, H.G. Robinson, L. Hollberg, ``Optical-pumping noise in laser-pumped, all-optical microwave frequency references", J. Opt. Soc. Am. B \textbf{18}, 1676 (2001).

\bibitem{martinelli} M. Martinelli, P. Valente, H. Failache, D. Felinto, L.S. Cruz, P. Nussenzveig, A. Lezama, ``Noise spectroscopy of nonlinear magneto-optical resonances in Rb vapor", Phys. Rev. A \textbf{69}, 043809 (2004).

\bibitem{valente} P. Valente, H. Failache, A. Lezama, ``Diode laser noise-spectroscopy of low-frequency atomic fluctuations in rubidium vapor", Eur. Phys. J. D \textbf{50}, 133 (2008). 

\bibitem{zibrov} A.S. Zibrov, I. Novikova, ``Observation of quantum noise in the polarization of laser light in a rubidium-vapor cell", JETP Lett. \textbf{82}, 110 (2005).

\bibitem{yashchuk} V.V. Yashchuk, D. Budker, J.R. Davis, ``Laser frequency stabilization using linear magneto-optics", Review of Scientific Instruments \textbf{71}, pp.341-346 (2000).

\bibitem{aleksanyan} A. Aleksanyan, S. Shmavonyan, E. Gazazyan, A. Khanbekyan, H. Azizbekyan, M. Movsisyan, A. Papoyan, ``Fluorescence of rubidium vapor in a transient interaction regime", J. Opt. Soc. Am. B \textbf{37}, 203 (2020).

\bibitem{steck} D.A. Steck, ``Rubidium 85 D line data", 01 2015 [Online]. Available: \url{https://steck.us/alkalidata}.

\bibitem{sarkisyan} D. Sarkisyan, T. Varzhapetyan, A. Sarkisyan, Yu. Malakyan, A. Papoyan, A. Lezama, D. Bloch, M. Ducloy, ``Spectroscopy in an extremely thin vapor cell: Comparing the cell-length dependence in fluorescence and in absorption techniques", Phys. Rev. A \textbf{69}, 065802 (2004).

\end{thebibliography}
\end{document}